\documentclass[preprint,showpacs,nofootinbib]{revtex4}
\usepackage{graphicx}
\begin{document}

\title{Charge distribution and radii in clusters from nuclear pasta models}

\author{M. \'{A}ngeles P\'{e}rez-Garc\'{i}a~\footnote{mperezga@usal.es}}
\affiliation{Department of Fundamental Physics and IUFFyM, \\University of Salamanca, 
Plaza de la Merced s/n 37008 Salamanca}

\date{\today}

\begin{abstract}
We study the consistency of the description of charge distributions and radii of nuclear clusters obtained with semiclassical nuclear pasta models. These nuclei are expected to exist in the low density outer crust of neutron stars. Properties of the arising clusterized nucleon matter can be compared to realistic nuclear properties as experimentally extracted on earth. We focus on non iso-symmetric light clusters with nucleon number $8 \le A \le 30$ and use Monte Carlo many-body techniques. We simulate isotopic chains for a set of selected nuclei using a model Hamiltonian consisting of the usual kinetic term, hadronic nucleon nucleon (NN), Coulomb and an effective density dependent Pauli potential. It is shown that for neutron rich (deficient) clusters neutron (proton) skins develop. Different (matter, neutron, proton, electric charge) radii are computed for this set of non iso-symmetric nuclei. Nuclear binding energies are also analyzed in the  isotopic chains.  
\end{abstract}
\vspace{1pc}

\maketitle

\section{Introduction}

The knowledge of nucleon distribution in  finite nuclear systems is a relevant topic in  nuclear physics  since it is not only important in characterizing the spatial extent of the nucleus but it is related to other issues like, for example, the location of the nucleon drip-lines~\cite{drip} or the determination of nuclear masses in the r-process nucleosynthesis in the astrophysical context~\cite{pfeiffer,tera}. A variety of approaches  have provided with nuclear mass formulae that allow testing properties of currently available nuclei from experiment~\cite{nucdata}. Both non relativistic  density-dependent Skyrme~\cite{skyrme} and Gogny interactions~\cite{gogny} and relativistic mean field (RMF) calculations~\cite{gambhir} have been largely used as they provide a successful reproduction of charge distribution and size of nuclei. Other many-body techniques like Monte Carlo, classical Molecular Dynamics, Antisymmetrized Molecular Dynamics or even more elaborate as the auxiliary field diffusion Monte Carlo have the potentiality of being applied to this type of calculations~\cite{kanada,ono,kanada2,afdmc,aka,aka2}. However, in the context of direct simulation of clusterized matter in systems such as the outer crust of neutron stars or supernova matter these techniques are not suitable.  Due to the nature of these intensive calculations it is usual to construct nuclear interaction models in a way so that they will retain the essential physics involved, yet they keep as simple as possible for saving CPU time. 

In exploratory works~\cite{aka,aka2} we used Monte Carlo techniques to study nuclear 
binding energies and nuclear density distributions for iso-symmetric spin saturated medium mass nuclei with nucleon number up to $A \approx 100$ using semiclassical simulations via Monte Carlo many-body techniques. In \cite{aka} the purpose was to study the role of an effective Pauli 
potential dependent on density  parameterized by one nuclear variable, the Fermi momentum of the constituent nucleons for nuclei with A up to $\approx 50$. In this semiclassical simulation of a finite nucleon system it was demonstrated that the empirical binding energies for these nuclei can be reproduced satisfactorily even if one uses a simplified NN interaction provided a density dependent Pauli potential is included. In that case a delicate counterbalance of kinetic and potential energy contributions arises allowing for the nuclear saturation mechanism. In particular, only isospin and additionally spin dependence in the NN interaction sector was included to size the impact on binding of (anti) alignement of nucleon spins. Obtained results show that this latter improvement in the description of the interaction weakly impacts the  reproduction of experimental nuclear binding energies.

In \cite{aka2} we considered a wider range of nuclei with $8 \le A \le 100$ 
for spin-isospin saturated $Z=N$ (even Z and N) nuclei. We showed that as the nucleon Fermi momentum is no longer a good parameter due to the fact that the average Fermi momenta in nuclei with $A > 50$ saturate to a value $\simeq 260$ MeV/c (e.g., $265$ MeV/c for a $^{208}$Pb nucleus)~\cite{moniz}, and additional dependence must be considered in order to parametrize fermionic correlations. This result, then, generalizes the previous conclusion that the density dependence of the Pauli potential is crucial to reproduce the empirical nuclear binding energies~\cite{aka}.
If heavier systems are explored, one obtains that for systems with nucleon number $A$ larger than $\approx 200$ the binding energies per particle and Pauli potential strength contribution stabilizes and the corresponding binding energy per particle for symmetric nuclear matter at low temperatures of $\approx -16$ MeV is obtained. 

As mentioned, in this type of semiclassical simulations one considers effective degrees of freedom under the form of nucleons. In this sense this treatment is appropriate to the interaction range of $fm$ (or energies of the order of the pion mass $m_\pi\approx 140$ MeV) in an analogous way to the very successful effective field theories (EFT) used in hadronic physics~\cite{QHD}. 

Other type of nuclear systems of interest in fields such as nuclear astrophysics can be considered in this simulation scenario. Clustered configurations arising in low density neutron star matter, below densities $\approx 0.1\rho_0$ ($\rho_0=0.17\,fm^{-3} $) and known as {\it pasta phase}~\cite{pasta, Watanabe, Maruyama, hor} can be studied using this approach, beyond the usually adopted Wigner-Seitz cell approximation that lacks the inclusion of higher order correlations~\cite{wigner}. In simulations of heavy-ion collisions the fragmentation yields at late times show that the final mass distribution of clusters is a relevant topic that largely depends on higher order correlations in the simulations~\cite{fragment}.

From the experimental point of view there is  an increasing effort in the experimentation with radioactive beams such as, among others, ISOLDE~\cite{isolde}, those at RIKEN~\cite{riken} or the proposed RIA~\cite{ria} and FAIR~\cite{fair}. Special attention is devoted to the new available (proton) neutron rich nuclei where observables such as the (proton) neutron skin radius can be measured. Experiments, like {\it e. g.} P-ReX at Hall A in Jlab, will measure the neutron radius in $^{208}Pb$ using  parity violating electron scattering~\cite{prex}. Additional phenomenology such as halo nuclei~\cite{Tanihata} has also been observed. 

The structure of the manuscript is as follows. In section~\ref{nuc} we present the effective nuclear model used in our simulations. In section~\ref{results} we present the results for the observables obtained from the calculations of the different nuclear isotopic chains considered in this work and finally we conclude in section~\ref{conclusion}.

\section{Nuclear simulation model}
\label{nuc}
In the present approach we describe matter in nuclear astrophysical scenarios treating nucleons as classical, structureless particles.  
The model Hamiltonian consists of nucleon kinetic energy, nuclear interaction ($V_{NN}$), 
Coulomb ($V_{Coul}$) and Pauli ($V_{Pauli}$) potentials.
The Pauli potential depends not only on relative positions but on momenta of the interacting nucleons. It simulates nucleon fermionic nature using a Gaussian functional form introduced by Dorso {\it et al.}~\cite{dorso} allowing, however, for density dependence, see below.

The model Hamiltonian is given by, 
\begin{equation}
H=\sum_{i=1}^{A} \frac{{\bf p}_{i}^{2}}{2m_N}
+ \sum_{i=1,j>i}^{A} 
\left[ {V}_{NN}(r_{ij})+V_{Coul}(r_{ij})
+V_{Pauli} (r_{ij},p_{ij}) \right], 
\label{ham}
\end{equation}
where ${\bf p}_{i}$ is the 3-momentum of $i$-th nucleon and 
$r_{ij}=|{\bf r}_i-{\bf r}_j|$ ($p_{ij}=|{\bf p}_i-{\bf p}_j|$) 
the relative distance (momentum) of the $i$-th and $j$-th nucleons.$m_N$ is the nucleon mass. We set $\hbar=c=1$. 

Since we are interested in the description of finite clusters we will use a simplified phenomenological NN potential  interaction, similar to those used in nuclear pasta models \cite{hor}, consisting of two Gaussians as follows,

\begin{equation}
V_{NN}(r_{ij})=a e^{-r_{ij}^2/\lambda_1^2}+[b+c \tau_i \tau_j] e^{-r_{ij}^2/\lambda_2^2}
\label{NNpot}
\end{equation}
where $\tau_i$ ($\tau_j$) is the isospin third-component of
$i$-th ($j$-th) nucleon ($+1$ for protons, $-1$ for neutrons). We will label this interaction potential as GM model. Let us emphasize that this model has a functional dependence that, although simple, can be used to model extended nuclear systems at high densities (several times nuclear saturation)~\cite{pasta}. It includes an isospin dependent part to ensure that while pure neutron matter is unbound, symmetric nuclear matter is appropriately bound \cite{hor}. However, in this work we are mainly interested in studying the low density limit of nuclear systems retaining the same functional form in the NN interaction. In this way a consistent check can be performed using the same treatment for nuclear systems in the whole density range where the nucleon picture holds. It has been shown, for example, that the lack of consistency between results for matter response to a weakly interacting probe, as obtained with fine tuned models for a limited range of systems, can lead to, for example, overpredict neutrino opacities in the astrophysical context~\cite{hor}. For accurate nuclear shell calculations one should relay, however, on more elaborate treatments including the before mentioned richer NN models with additional two and three-body forces as in Quantum Monte Carlo~\cite{shell} or {\it ab initio} calculations~\cite{abinitio}.
\begin{table}[htbp]
\begin{center}
\caption{Model parameters.}
\label{tab:param}
\begin{tabular}{lllll} \hline \hline
$a\,(MeV)$ &$b\,(MeV)$  &$c\,(MeV)$ & $\lambda_1\,(fm)$ &$\lambda_2\,(fm)$\\ \hline
$50$  &$-4$  &$3$ &$0.8$ &$4$ \\\hline \hline
\end{tabular}
\end{center}
\end{table}

In order to calibrate the model we choose a set of mirror nuclei where the neutron number, $N$, and the proton number, $Z$, can be interchanged as, for example, in $^{14}C$ and $^{14}O$ and fit the experimentally available root mean square (RMS) matter and charge radius. The parameter values used in this work are given in Table~\ref{tab:param}. Proton and charge radius can currently be measured with high relative precision using electromagnetic probes by the cross section measurement of elastic  scattering of electrons~\cite{oxy2} or muonic atoms~\cite{muon}. However, for matter and neutron radii the accuracy is not so good due to the fact that there is uncertainty in the nuclear regions probed by the various processes and due to the effective nuclear interactions between nucleons. Antiprotonic atoms~\cite{antip} provide information on neutron distribution in a nucleus.

Since protons are electrically charged particles we introduce the Coulomb potential,
\begin{equation}
V_{Coul}(r_{ij})=  
\frac{e^2}{4 \pi r_{ij}} \frac{1+\tau_i}{2} \frac{1+\tau_j}{2},   
\label{Coulombpot}
\end{equation}
where $e$ is the proton electric charge. 

The antisymmetrization is obtained in an effective way by using a Pauli potential under the Gaussian form given by,
\begin{equation}
{V}_{Pauli}({r}_{ij}, {p}_{ij})= 
V_P\,\,  e^{-\frac{r_{ij}^2}{2q_0^2}
-\frac{p_{ij}^2}{2p_0^2}} \delta_{\tau_i \tau_j}
\delta_{\sigma_i \sigma_j}, 
\label{Paulipot}
\end{equation}
where $\delta_{\tau_i \tau_j}$ ($\delta_{\sigma_i \sigma_j}$) is  
the Kronecker's delta for the isospin (spin) third-component. 
It prevents nucleons from occupying the same phase 
space volume when they have the same 
quantum numbers. We will allow a density dependence for this Pauli potential in the parameters $q_0,p_0$ and $V_P$ as in ~\cite{aka,aka2}. Although this effective treatment using a Pauli potential is a way of providing antisymmetrization it has some well known shortcomings~\cite{newpauli}.

In our simulation a nucleus is considered to be a composite object where $A$ non-relativistic point-like nucleons aggregate according to the interaction given by the Hamiltonian in Eq.(\ref{ham}). The typical internucleon distance in an iso-symmetric nucleus is given by $\approx 2r=\left(\frac{6}{\pi \rho}\right)^{1/3}$ where $\rho=2 p_F^3/3 \pi^2$ is the nucleon density and $p_F$ the nucleon Fermi momentum. In Moniz {\it et al.}~\cite{moniz} Fermi momentum is  asignated by using a simplified Fermi gas model to parameterize the density dependence of electron scattering cross section on a set of selected nuclei. To study the structure of the clustered configurations arising in the simulation one can compute the RMS radius by averaging the corresponding density operator over the simulation box volume ($V=L^3$).

In this work we will consider different charge (matter, neutron, proton) density operators that will be designated as $\rho_i$ with $i=\{m, n, p\}$ and charge $q_i$ with $i=\{1,1/2-\tau_j,1/2+\tau_j\}$ respectively and $j=1,..,A$. The charge density operators can be written generically as
\begin{equation}
\rho_i({\bf r})=\frac{1}{L^3} \sum_{j=1}^{A} q_i \delta({\bf r}-{\bf r_j})
\end{equation}
Let us note that the RMS radius is computed by averaging the corresponding charge operator over the box volume. In this way the RMS radius can be obtained as 
\begin{equation}
<r_i^2>^{1/2} =\left[ \frac{\int d^3 {\bf r} \rho_{i}({\bf r})\,r^2 }{\int d^3 {\bf r} \rho_{i}({\bf r}) }\right]^{1/2}
\label{ri}
\end{equation}
in order to compare with the nuclear charge radius, $r_{ch}$, that can be experimentally measured, it is usual~\cite{buchin} to assume a Gaussian charge distribution inside the proton so that an effective proton finite size, $s_p$, is used and typically $s_p \approx 0.8$ fm~\cite{proton}. The expresion that will be taken for the charge RMS radius is, 
\begin{equation}
<r_{ch}^2>^{1/2} =[<r_p^2>+s_p^2]^{1/2}
\end{equation}
where $<r_p^2>^{1/2}$ is the point proton charge RMS radius.

Having explicited the different RMS radii above we can proceed for neutron rich (deficient) clusters and define the difference of RMS radii of the neutron and proton distribution \cite{del}, $\Delta r_{np}$, as
\begin{equation}
\Delta r_{np}=<r_n^2>^{1/2}-<r_p^2>^{1/2}
\label{rns}
\end{equation}

In a similar way we define $\Delta r_{pn}=-\Delta r_{np}$ for the proton rich clusters. There are estimates of this observable in the literature such as in~\cite{myers} where they quote the droplet model based value $\Delta RMS=\frac{3}{5}t$ where $t$ is the neutron skin thickness or the shell model potential extimates \cite{wesol}. Note that diffuseness of the surface and non-uniformity of densities add corrections to this last expression. This observable is also closely related~\cite{ratio} to the shift in the surface to volume ratio of symmetry energy, $r_{S/V}=a_S/a_V$ as derived in Skyrme-Hartree-Fock models~\cite{SHF} and other works~\cite{rsv2, formula}.

In \cite{fit} a linear fit is given using a set of experimental data measured in antiprotonic atoms as the isospin asymmetry parameter $I=(N-Z)/A$ variates. They give an expression for the fit of the neutron skin  
\begin{equation}
\Delta r_{np}= (0.9\pm 0.15)I+(-0.03\pm 0.02)\,fm
\label{wes}
\end{equation}
Notice however that this is just a fit to medium size nuclei. For smaller A there are also some calculations such as~\cite{itagaki,lala} where they consider the light nuclear sector. Additional information using this data can be derived for the symmetry energy~\cite{centelles}. 

\section{Results}
\label{results}
In the simulation we initially place the finite nuclear system under study, consisting of $A$ nucleons  distributed uniformly inside a sphere of radius $R_0$ in the range $2-3$ fm, within a cubic box of volume $V=L^3$ and impose $L >> R_0$.
Computer simulations are performed by Monte Carlo techniques where using the Metropolis algorithm~\cite{Metropolis} the ground state configuration is searched by thermal relaxation at a fixed temperature $T=1$ MeV. Note that the role played by  temperature in this semiclassical approach is just that of a parameter to somewhat simulate the effect of zero-point motion in a phenomenological way.
Once the thermalized stable clustered configuration is obtained, a nucleus, we statistically sample the system for the averaged magnitudes, namely, binding energies and RMS radii.

\begin{figure}[hbtp]
\begin{center}
\includegraphics [angle=-90,scale=0.75] {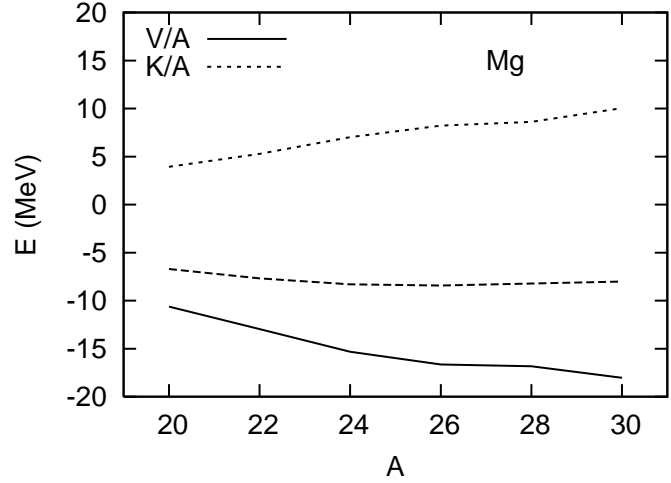}
\caption{Potential (solid line), kinetic (dotted line) and total (dashed line) binding energy per nucleon for the isotopic chain of $Mg$. 
}
\label{fig-n10}
\end{center}
\end{figure}

In Fig.~\ref{fig-n10} we can see the energetic contributions to the Mg isotope chain~\cite{ren}. Solid, dotted and  dashed lines show the potential, kinetic and total binding energy, respectively. Positive (kinetic) and negative (potential) energetic contributions allow to reproduce the experimental isotopic binding energy values. Note that due to the Pauli potential the kinetic energy per particle departs from the value $k=\frac{3}{2} k_B T$ as one would expect in an uncorrelated classical plasma.

\begin{figure}[hbtp]
\begin{center}
\includegraphics [angle=-90,scale=0.75] {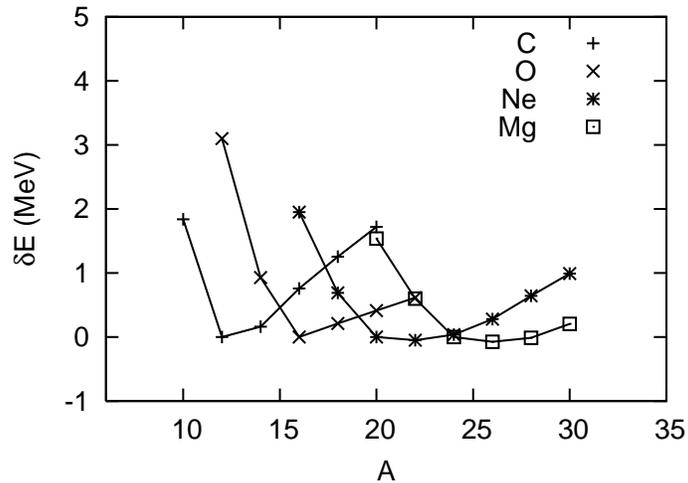}
\caption{Energy shift per nucleon with respecto to the $N=Z$ species  for the isotopic chains of C, O, Ne and Mg nuclei. }
\label{fig-n1}
\end{center}
\end{figure}

In Fig.~\ref{fig-n1} we show the shift in binding energy per nucleon 
($\delta E$) from the spin saturated iso-symmetric species ($N=Z$) versus nucleon number $A$ for Carbon (C), Oxygen (O), Neon (Ne) and Magnesium (Mg) isotopes as obtained by our simulation. We can see that for Ne and Mg the minimum is shifted to the neutron rich side in agreement with experimental binding energy data \cite{nucdata}.

\begin{figure}[hbtp]
\begin{center}
\includegraphics [angle=-90,scale=0.75] {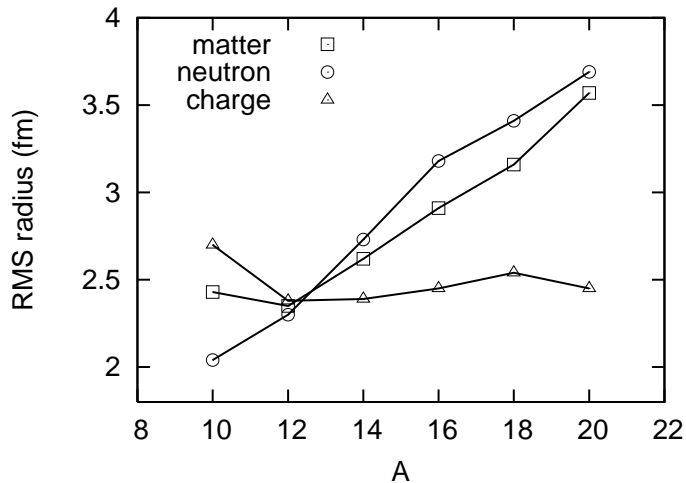}
\caption{From top to bottom: neutron, matter and charge RMS radii for the isotopic chain of Carbon.
}
\label{fig-n2}
\end{center}
\end{figure}

In Fig.~\ref{fig-n2} we plot  matter (boxes), neutron (circles) and charge (triangles) RMS radii for the Carbon isotopic chain versus the nucleon number $A$. We can see that for the iso-symmetric species with $Z=6$,  $N=6$ there is no skin present. As $A$ increases this neutron skin grows since the charge radius remains approximately constant. For the proton rich isotopes a proton skin develops while for nuclei on the neutron rich side a neutron skin forms in agreement with more elaborated relativistic mean field~\cite{sharma} or relativistic Hartree-Bogoliuvov \cite{lala} descriptions including nuclear deformations. 

\begin{figure}[hbtp]
\begin{center}
\includegraphics [angle=-90,scale=.75] {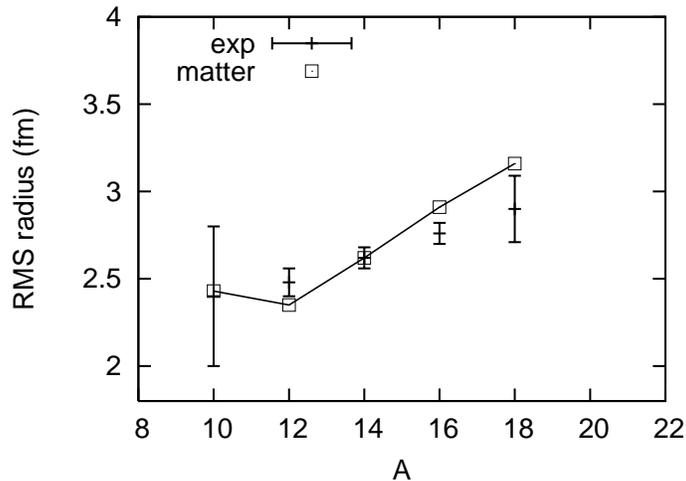}
\caption{Matter RMS radii for the isotopic chain of Carbon. Empirical data are shown along with error bars taken from Ref.~\cite{ozawa}.
}
\label{fig-n3}
\end{center}
\end{figure}

We can see more in detail in Fig.~\ref{fig-n3} matter (boxes) RMS radii for the Carbon isotopes along with experimentally deduced values from Ref.~\cite{ozawa}. Data are extracted from hadronic reactions involving break up of nuclei. For example, in the reaction
\begin{equation}
a + A \rightarrow b + X
\end{equation}
the width of the momentum distribution, $\sigma$, depends on the mass number of the projectile ($A_a$) and of the fragment ($A_b$) but it is found to be essentially independent of the target mass and beam energy. It can be expressed using the Goldhaber model~\cite{goldhaber} as
\begin{equation}
\sigma=\sqrt{A_b (A_a-A_b)/(A_a-1)}\sigma_0
\end{equation}
with this prescription and in the approximation where each isotopic component is considered as a Fermi gas with Fermi momentum $p_F$, then the nucleon density is $\rho=\frac{2 p_F^3}{3 \pi^2}$ and $\sigma_0$ is the reduced width of the momentum distribution and it is associated with the Fermi momentum of the nucleons, $\sigma_0=p_F/\sqrt{5}$. The interaction cross section, $\sigma_I$, can be experimentally measured with $5\%$ accuracy and functionally can be parameterized as
\begin{equation}
\sigma_I=\pi (R_I +r_0 A^{1/3})^2
\end{equation}
$R_I$ is the interaction radius for the selected projectile on the target nucleus. Fitting correspondingly $r_0$ to the experimental cross section data one obtains the general trend for the interaction cross section. For neutron rich isotopes there is, however, a deviation from this prediction as a neutron skin forms.
 Using Glauber model calculations the matter RMS radius can be estimated by using the approximate relation above~\cite{Tanihata}.

\begin{figure}[hbtp]
\begin{center}
\includegraphics [angle=-90,scale=.75] {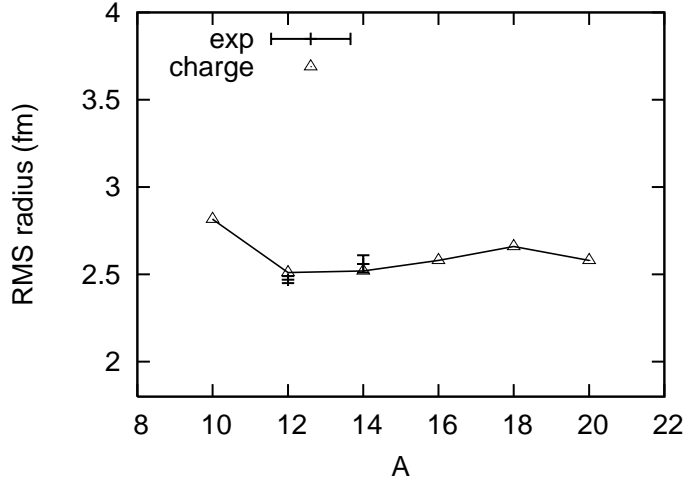}
\caption{Same as Fig.~\ref{fig-n3} for the charge  RMS radii.
}
\label{fig-n4}
\end{center}
\end{figure}

\begin{figure}[hbtp]
\begin{center}
\includegraphics [angle=-90,scale=.75] {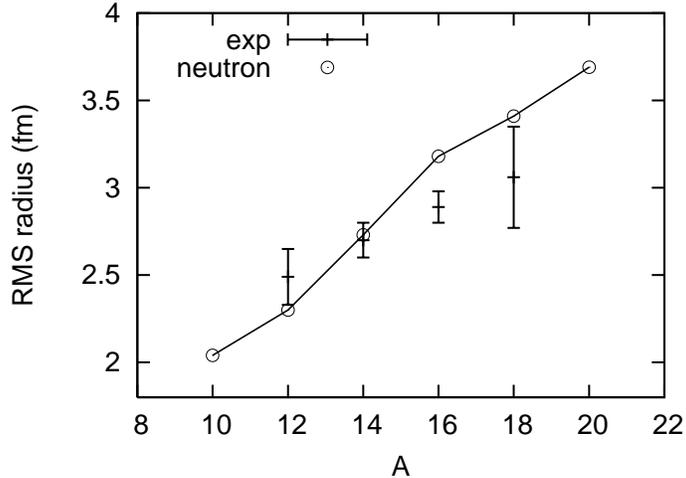}
\caption{Same as Fig.~\ref{fig-n3} for the neutron  RMS radii.
}
\label{fig-n5}
\end{center}
\end{figure}

In Figs.~\ref{fig-n4} and ~\ref{fig-n5} we plot  charge and neutron RMS radii, respectively,  for the Carbon isotopes along with experimental data. We can see that while the neutron RMS radius is a rising function of $A$ the nuclear charge radius stabilizes from the iso-symmetric isotope ($T_z=0$) up to values $T_z=4$ where $T_z=(N-Z)/2$ is the third-component of the isospin. Note that in our analysis we do not consider deformated nuclei since a many-body $r$-dependent potential used in nuclear pasta calculations is used. Deformations, $\beta_2$, in the carbon isotopic chain range from $[-0.4, 0.4]$ \cite{lala}. 

\begin{figure}[hbtp]
\begin{center}
\includegraphics [angle=-90,scale=0.75] {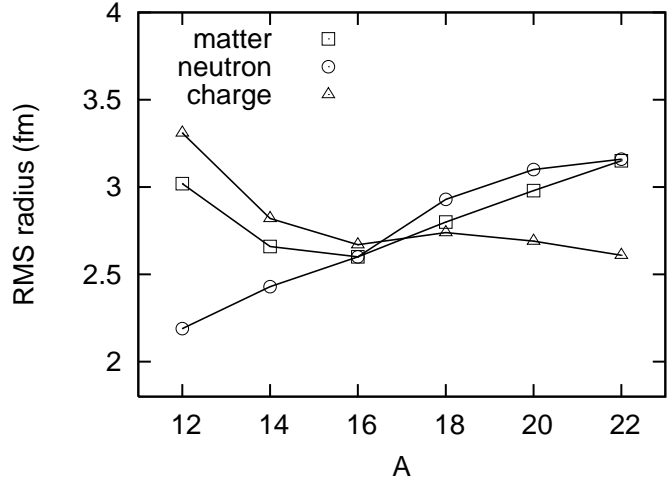}
\caption{Matter (boxes), neutron (circles) and charge (triangles) RMS radii for Oxygen isotopes.
}
\label{fig-n6}
\end{center}
\end{figure}

In Fig.~\ref{fig-n6} we show matter RMS (boxes), neutron (circles) and charge (triangles) RMS radii for the Oxygen isotopes. We can see that as isospin asymmetry grows a neutron (proton) skin develops on the neutron (proton) rich side, that is thinner than in the case of Carbon as shown in Fig.~\ref{fig-n2}.

\begin{figure}[hbtp]
\begin{center}
\includegraphics [angle=-90,scale=0.75] {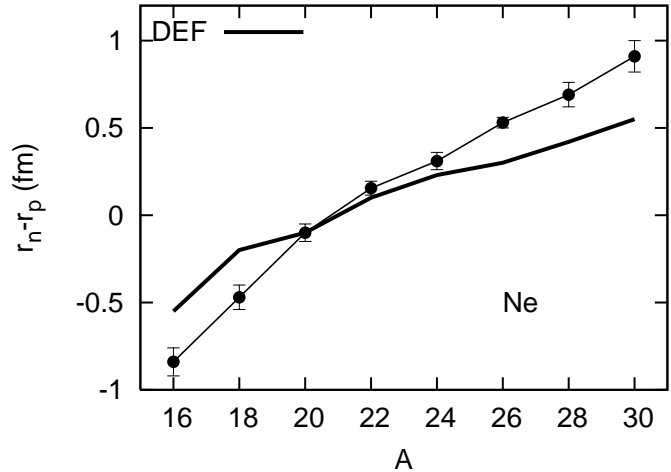}
\caption{Difference in proton and neturon RMS radii versus nucleon number $A$ for the Neon isotope chain as deduced from the DEF model~\cite{neonrad} and compared to our calculation with the GM model.
}
\label{fig-n9}
\end{center}
\end{figure}

In Fig.~\ref{fig-n9} we can see the difference in neutron and proton RMS radii, $\Delta_{np}$, as defined in Eq.(\ref{rns}), versus nucleon number $A$ for the Neon isotope chain, as obtained in more sofisticated calculations as those of the RMF approximation with the constant gap approximation in the axially deformed oscillator basis (DEF) model from Ref.~\cite{neonrad} and compared to our semiclassical simulation with the GM model. We can see that the divergence of both models is bigger as the isospin asymmetry grows. However if we consider the result from the linear fit in Eq. (\ref{wes}) at the highest isospin asymmetry considered in this chain, at A=30, the discrepancy is about $16\%$. This shows that the description provided by a simplified nuclear interaction, as that of the GM model, similar to those used in nuclear pasta models allows a reasonable consistent description as compared to more elaborate treatments. 

\begin{figure}[hbtp]
\begin{center}
\includegraphics [angle=-90,scale=0.75] {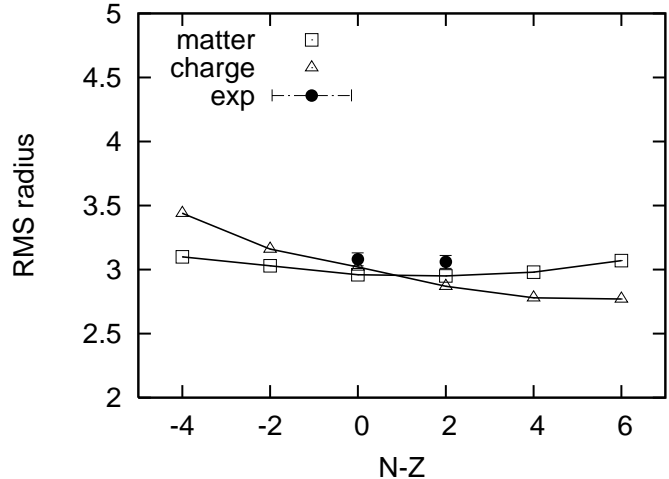}
\caption{Matter and charge RMS radii versus the neutron excess for the $Mg$ isotopic chain. Experimental values for matter radii are depicted for comparison as taken from Ref.~\cite{ozawa}.
}
\label{fig-n11}
\end{center}
\end{figure}

In Fig.~\ref{fig-n11} we show matter and charge RMS radii versus neutron excess for the Mg isotopic chain. Experimental values for matter radii are depicted for comparison as taken from Ref.~\cite{ozawa}. Neutron excess (fault) in the isotope chain results in a developing neutron (proton) skin thinner than in the case of Carbon and Oxygen isotopes. As we can see there is a competing effect that tends to enlarge the neutron skin as the isospin asymmetry grows and that tends to decrease the neutron skin as the atomic number incresases. 

\section{Conclusions}
\label{conclusion}
In this work we anlyze the consistency of the description of finite nuclear systems, clusters, in the limit of low density of interest in the astrophysically relevant nuclear {\it pasta} models. In this regime densities in the range of tenths to several times nuclear saturation density are considered. We obtain some observable properties of finite nuclei regarding charge distribution and different (matter, neutron, proton, charge) RMS radii for spin saturated nuclei in the mass range with $8 \le A \le 30$. We use many-body semiclassical simulations with Monte Carlo techniques to compare the results provided by such a nuclear interaction, designed to be used in many-body systems to experimental data or more elaborated nuclear models. 

We use a hamiltonian model where apart from the kinetic term, the NN interaction consists of a hadronic, Coulomb and Pauli effective potential terms. We have shown that for a set of nuclear isotopic chains (C, O, Ne, Mg) experimental binding energies can be well reproduced if a density dependence for the Pauli potential is allowed. We have also analyzed other  nuclear structure properties such as matter, neutron and charge radii. We have found, besides, that our obtained radii agree reasonably well with those deduced from experiment. As the isospin asymmetry parameter $T_z$ grows, the departure from this trend, as obtained with our simplified model, is larger remaining below $16\%$ however.

For non iso-symmetric clusters we find that a neutron (proton) excess results in a neutron (proton) skin developing in the exterior shell of the nuclei and computed as the difference of neutron and proton RMS radii. For iso-symmetric nuclei in our model there is a vanishing skin, in accordance with the liquid drop formula.
Charge radii are computed in our simulation from point-like particles by folding the proton RMS radius with the proton finite size $s_p \approx 0.8$ fm. Our computed values agree with measured values within experimental error bars. Charge radii stabilizes as A increases giving the possibility of neutron skin formation on the neutron rich sector. However, in particular, the Neon isotope anomaly~\cite{neonrad} can not be reproduced within our simplified spherical configuration. For this type of phenomenology nuclear deformations play an important role. Additional enrichment of the model using two and three-body terms could provide better agreement.
Using these techniques of semiclassical simulations on effective degrees of freedom allows a consistent procedure to further testing asymmetric systems by studying the formation of clusters in these environments and how their structure affects the opacity of this type of matter.

\vspace{2ex}
\noindent{\bf Acknowledgments}\\
We would like to thank useful discussions with K. Tsushima and A. Valcarce and their interest in this work. Partial financial support  from the Spanish Ministry of Education and Science project DGI-FIS2006-05319 is gratefully acknowledged.

\end{document}